\documentclass[sn-mathphys-num]{sn-jnl}% Math and Physical Sciences Numbered Reference Style 
\usepackage{graphicx}%
\usepackage{caption}%
\usepackage{multirow}%
\usepackage{amsmath,amssymb,amsfonts}%
\usepackage{amsthm}%
\usepackage{mathrsfs}%
\usepackage[title]{appendix}%
\usepackage{xcolor}%
\usepackage{textcomp}%
\usepackage{manyfoot}%
\usepackage{booktabs}%
\usepackage{algorithm}%
\usepackage{algorithmicx}%
\usepackage{algpseudocode}%
\usepackage{listings}%
\begin{document}

\title[Article Title]{GPU-based data processing for speeding-up correlation plenoptic imaging}


\author*[1]{\fnm{Francesca} \sur{Santoro}}\email{santoro@planetek.it}
\author[1]{\fnm{Isabella} \sur{Petrelli}}%\email{petrelli@planetek.it}
\author*[2,3]{\fnm{Gianlorenzo} \sur{Massaro}}\email{gianlorenzo.massaro@uniba.it}
\author[4]{\fnm{George} \sur{Filios}}%\email{filios@planetek.gr}
\author[2,3]{\fnm{Francesco V.} \sur{Pepe}}%\email{francesco.pepe@ba.infn.it}
\author[1]{\fnm{Leonardo} \sur{Amoruso}}%\email{amoruso@plnaetek.it}
\author[4]{\fnm{Maria} \sur{Ieronymaki}}%\email{ieronymaki@planetek.gr}
\author[5]{\fnm{Samuel} \sur{Burri}}%\email{samuel.burri@epfl.ch}
\author[5]{\fnm{Edoardo} \sur{Charbon}}%\email{edoardo.charbon@epfl.ch}
\author[5]{\fnm{Paul} \sur{Mos}}%\email{paul.mos@epfl.ch}
\author[5]{\fnm{Arin} \sur{Ulku}}%\email{arin.ulku@epfl.ch}
\author[5]{\fnm{Michael} \sur{Wayne}}%\email{michael.wayne@epfl.ch}
\author[1]{\fnm{Cristoforo} \sur{Abbattista}}%\email{abbattista@planetek.it}
\equalcont{These authors contributed equally to this work.}
\author[5]{\fnm{Claudio} \sur{Bruschini}}%\email{claudio.bruschini@epfl.ch}
\equalcont{These authors contributed equally to this work.}
\author[2,3]{\fnm{Milena} \sur{D'Angelo}}%\email{milena.dangelo@uniba.it}
\equalcont{These authors jointly supervised this work.}

\affil[1]{\orgname{Planetek Italia s.r.l.}, \orgaddress{\city{Bari}, \postcode{I-70132}, \country{Italy}}}
\affil[2]{\orgname{Dipartimento Interateneo di Fisica, Universit\`a degli Studi di Bari}, \orgaddress{\city{Bari}, \postcode{I-70125}, \country{Italy}}}
\affil[3]{\orgname{INFN - Sezione di Bari}, \orgaddress{\city{Bari}, \postcode{I-70125}, \country{Italy}}}
\affil[4]{\orgname{Planetek Hellas E.P.E.}, \orgaddress{\city{Athens}, \postcode{15125}, \state{Marousi}, \country{Greece}}}
\affil[5]{\orgname{\'Ecole polytechnique f\'ed\'erale de Lausanne (EPFL)}, \orgaddress{\city{Neuch\^{a}tel}, \postcode{2002}, \country{Switzerland}}}

%%==================================%%
%%    Abstract                      %%
%%==================================%%
\abstract{Correlation Plenoptic Imaging (CPI) is a novel technological imaging modality enabling to overcome drawbacks of standard plenoptic devices, while preserving their advantages. However, a major challenge in view of real-time application of CPI is related with the relevant amount of required frames and the consequent computational-intensive processing algorithm.
In this work, we describe the design and implementation of an optimized processing algorithm that is portable to an efficient computational environment and exploits the highly parallel algorithm offered by GPUs. Improvements by a factor ranging from 20x, for correlation measurement, to 500x, for refocusing, are demonstrated. Exploration of the relation between the improvement in performance achieved and actual GPU capabilities, also indicates the feasibility of near-real time  processing capability, opening up to the potential use of CPI for practical real-time application.}



\maketitle

%%==================================%%
%%    Introduction                  %%
%%==================================%%
\section{Introduction}\label{sec1}

Correlation Plenoptic Imaging (CPI) is a novel imaging modality exploiting spatio-temporal intensity correlations of light to enable plenoptic (or light-field) imaging at the diffraction limit \cite{dangelo2016correlation,pepe2016plenoptic,pepe2016correlation,pepe2017exploring,pepe2017diffraction,massaro2021lightfield,abbattista2021towards,massaro2023}. In traditional plenoptic imaging techniques, based on direct intensity measurement, information on light spatial distribution and direction are encoded in a single detector \cite{lippmann1908epreuves,adelson1992single,ng2005light}, thus entailing spatial resolution reduction. Based on the different physical mechanism that underlies the standard and the correlation approach to light-field imaging, CPI considerably extends the accessible volumetric resolution \cite{pepe2017diffraction,massaro2024assessing,scattarella2022resolution,Scattarella2023}.
Several alternative configurations of CPI have so far been proposed \cite{pepe2017exploring,scagliola2020correlation,dilena2018correlation,dilena2020correlation,abbattista2021towards}, working with the intrinsic correlations of either chaotic light \cite{dangelo2016correlation} or entangled photon pairs \cite{pepe2016correlation}.
However, measuring light spatio-temporal correlations imposes the typical drawback of correlation imaging, namely the need to  collect a statistically relevant number of frames to reconstruct the pixel-by-pixel correlation function with acceptable signal-to-noise ratio \cite{massaro2022comparative}; this potentially limits the acquisition and elaboration speed, imposing a challenge in view of real time applications (see, e.g., \cite{cassano2017spatial,dangelo2017characterization, pepe2022distance}). 

In Ref. \cite{massaro2023}, correlation photon imaging at a rate of 10 \textit{volumetric} images per second was achieved, thanks to the combination of chaotic light and the use of ultra-fast detectors: SwissSPAD2, an array of single photon avalanche photodiodes (SPAD) capable of capturing up to $10^5$ binary frames per second with ns resolution, while combining an extremely low noise and high detection efficiency \cite{ulku2019spad,ulku2020spad, antolovic2016photon, abbattista2021towards}. However, even if the overall measurement time benefits from the speed of the SPAD arrays, the binary nature of the acquired frames imposes the need to acquire an even larger number of frames with respect to previous CMOS-based CPI experiments \cite{massaro2023}, leading to a critical extension of the computational time \cite{s22072778,massaro2022refoc}. To address this challenge, in this article, we explore the adoption of ultra-fast computing platforms based on Graphic Processing Units (GPUs), and demonstrate a 20x to 500x improvement in elaboration time over the conventional CPU-based approach so far employed.
 
The paper is organized as follows: in Section \ref{sec2}, we provide a description of the sample dataset used for benchmarking the optimized CPI software (SW) version exploiting GPU. In Section \ref{sec3}, we present the software architecture developed in the frame of this activity as well as some specific software development kit (SDK) functionalities made available to facilitate CPI data management. In Section \ref{sec4}, we present the results obtained by optimizing the CPI algorithm for GPU-related environment, and in Section \ref{concl}, we discuss the conclusions and future perspectives.

%%==================================%%
%%    Input Data                    %%
%%==================================%%
\section{Input Dataset}\label{sec2}
\begin{figure}[tb] 
\centering 
\includegraphics[width=0.9\linewidth]{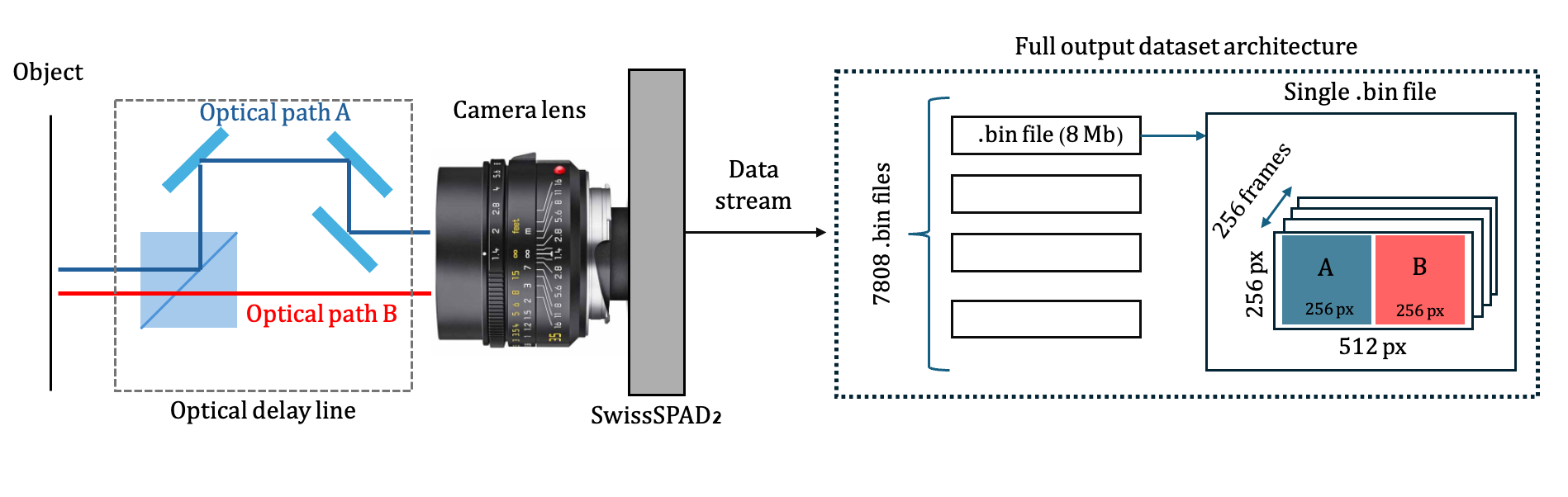}
\caption{Experimental setup and acquired data architecture. Further details about the experiment are available in Ref.~\cite{massaro2023}.}
\label{fig:setup} 
\end{figure}

\begin{figure}[tb] 
\centering 
\includegraphics[width=0.7\linewidth]{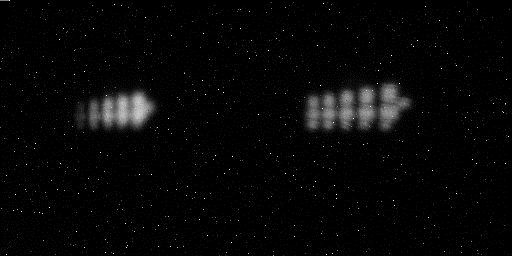}
\caption{Input Data preview: grey-scale image resulting from the cumulative sum of 512 sample binary frames acquired during the experiment using SwissSPAD2.}
\label{fig:inputData} 
\end{figure}

The GPU-optimized software solution presented in this work assumes that the relevant amount of input data needed for CPI processing has been already acquired and stored on file-system, formatted as binary data. The adopted data format is in line with the output interface (I/F) foreseen by SwissSPAD2 acquisition sensor boards \cite{ulku2019spad}, in order to facilitate the future perspective of concatenating easily the data acquisition and the data processing chains. The sample dataset used in our experiments, as shown in Fig.~\ref{fig:setup}, is detailed as follows: 
\begin{itemize}
  \item Dataset(s):
  \begin{itemize}
    \item \textit{Full Dataset}: Set of 7808 .bin files, 8MiB each
    \item \textit{Subset:} Set of 20 .bin files, 8MiB each
  \end{itemize}
  \item Each file contains a fixed amount of frames $N_\text{file}$ (i.e. 512 binary frames)
  \item Frame size: $256 \times 512$ px
  \item Bit-depth = 1 bit
\end{itemize}
Each of the $N_\text{file}$ frames within a .bit file is composed of two $256\times 256$ side-by-side images, as shown in Fig.~\ref{fig:inputData}, which shows an overview (cumulative sum of binary frames) of the used dataset.
Hence, the leftmost part of each frame (from pixel 1 to pixel 256) represents one image $A^i_{j\,k}$, and the rightmost part (pixel 257 to 512) represents a second image $B^i_{j\,k}$, where the index $i=1,\dots,N_\text{files}$ identifies the frame within the .bin file, and $j,k=1,\dots,256$ represent pixels coordinates along the horizontal and vertical, respectively.
By considering that the dataset consists of many .bin files ($N$), the total number of images $A$ and $B$ is $N_\text{TOT}=N\times N_\text{files}$.

Since our work only deals with the computational aspects related to the analysis of such datasets, further experimental details (\textit{e.g.} optical design, imaging parameters, noise performance) will not be reported.
Detailed discussion about the experimental aspects can be found in Ref. \cite{massaro2023}.

% The following tables include the configuration adopted for the experiment, \textit{i.e.} details about object target, light source (Table \ref{tabData1}), sensor (Table \ref{tabData2}) and camera (Table \ref{tabData3}) parameters setting used during data acquisition, while Figure \ref{fig:inputData} shows an overview (cumulative sum of binary frames) of the used dataset.

\section{CPI algorithm}
The algorithm for extracting the 3D stack of axial images from the dataset consists of two main blocks:
\begin{enumerate}
    \item \textbf{Evaluation of the correlation function.}
    A four-dimensional correlation function $\Gamma$ is evaluated from the $N_\text{TOT}$ images $A$ and $B$. $\Gamma$ is obtained by averaging the tensor product between corresponding images $A^i\otimes B^i$, and subtracting the tensor product of the two average images from that quantity:
    \begin{equation}
        \Gamma_{j\,k\,m\,n}=\frac 1{N_\text{TOT}}
        \sum_{i=1}^{N_\text{TOT}}
        A^i_{j\,m}\,B^i_{k\,n}-
        \left(\frac 1{N_\text{TOT}}
        \sum_{i=1}^{N_\text{TOT}}
        A^i_{j\,m}
        \right)
        \left(\frac 1{N_\text{TOT}}
        \sum_{i=1}^{N_\text{TOT}}
        B^i_{k\,n}
        \right).
        \label{eq:corrFunc}
    \end{equation}
    The computational cost of this process depends on the pixel size of the images and on the number of images;
    \item \textbf{Refocusing.}
    The 4D correlation function is then processed through an algorithm based on Radon transformations \cite{massaro2024assessing,massaro2022refoc}, which extracts the 3D stack of axial sections.
    Since it is applied after averaging over the number of planes, the computational requirements of this step only depend on the pixel size of the correlation function \cite{s22072778}.
    \end{enumerate}

%%==================================%%
%%    SW Architecture               %%
%%==================================%%
\section{Software Architecture}\label{sec3}
\subsection{Overview}\label{subsec31}
\begin{figure}[tb] 
\centering 
\includegraphics[width=0.9\linewidth]{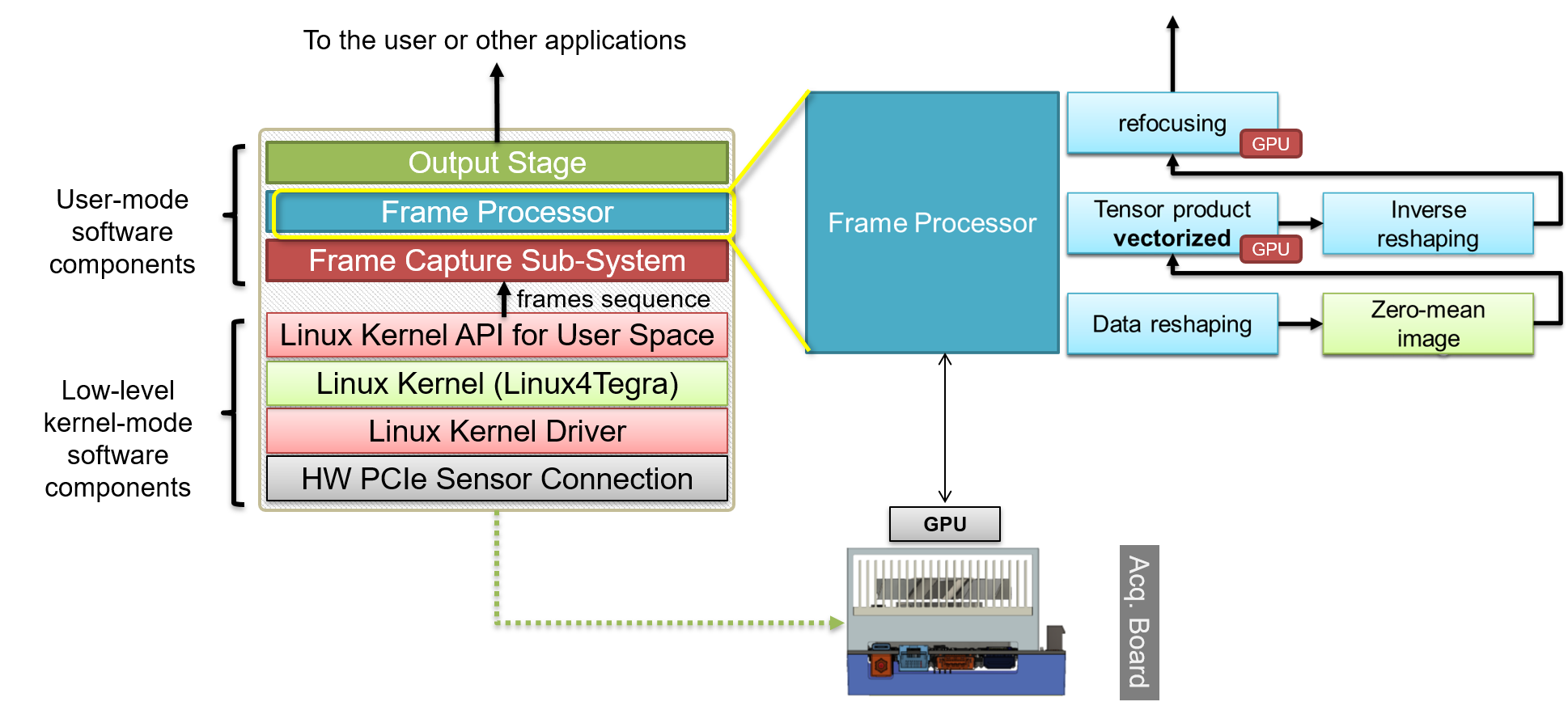}
\caption{High-level SW Architecture for CPI applications based on GPU acceleration. }
\label{fig:SWarch1} 
\end{figure}

The development of our optimized SW solution relies on a pre-existing proprietary software framework, designed and developed within the Planetek group, whose functionalities were customized to respond to requirements of improving CPI timing performances in a GPU-based environment.
The CPI software architecture depicted in Fig.~\ref{fig:SWarch1} is focused on the exploitation of a dedicated GPU-based acceleration device, and can be decomposed into two main categories of SW components: 
\begin{enumerate}
 \item \textbf{Low-level kernel-mode} software components, aimed at direct interfacing the HW setup responsible for data acquisition from sensor boards; it can be further divided into:
    \begin{enumerate}
     \item HW PCIe Sensor Connection
     \item Linux Kernel Driver
     \item Linux Kernel
     \item Linux Kernel API for User Space
    \end{enumerate}
 \item \textbf{User-mode} software components, responsible for the actual algorithmic processing and data delivery; it consists of:
    \begin{enumerate}
     \item Frame Capture Sub-System
     \item Frame Processor
     \item Output Stage
    \end{enumerate}
\end{enumerate}

A detailed overview of the Frame Processor SW component is then provided in Fig.~\ref{fig:SWarch1} (top-right), in which main steps of Correlation Plenoptic Imaging (CPI) algorithmic workflow are listed in sequence, highlighting those one identified as characterized by critical performances, and thus presented as good candidates for a GPU-based implementation. 

%%==================================%%
\subsection{Software Development Kit (SDK)}\label{subsec32}
The software framework considered as the baseline for this activity provides a multi-platform C++17 \cite{ISO:2017:IIP} development environment (an SDK) for the creation of software applications ranging from console-based data processors to interactive graphics tools, mainly dedicated to image processing and visualisation.
Its main goal is to provide a cross-platform development environment and runtime for creating general purpose applications that can run on desktop, server and embedded systems with a single code base. It achieves this goal by combining a rich set of dedicated classes with the functionality provided by the C++17 runtime library. 
As a framework, it is composed of a collection of dynamic libraries, and their respective include files, organized according to a hierarchical approach, based on the following three foundation libraries: 
\begin{itemize}
 \item \texttt{CORE}, which is responsible for abstracting the underlying operating system services for the upper layers of any application, so that access to the filesystem, some inter-process communication facilities, and other low-level services become cross-platform the user of the framework;  
 \item \texttt{MATH}, which provides basic mathematics functions and structures, such as vectors, matrices, simple algorithms; 
 \item \texttt{ACC}, which provides image manipulation and processing capabilities, abstraction of acceleration facilities (such as those based on OpenCL), and implementations of thread schedulers for multi-processing architectures. 
\end{itemize} 
	
On top of these three libraries, we have created other libraries with further specialisations and a rich ecosystem of plugins that extend the capabilities of the overall framework in terms of:
\begin{enumerate}
 \item Support for 3rd party image formats (such as jpeg, tiff, png, hdf5, etc.) both in reading and in writing; 
 \item Support for more algorithms, both of general usage and very specialized, encapsulated into entities named \textit{engines}, i.e. a special-purpose plugins; 
 \item Support for more acceleration devices and APIs. In this context, for example, the OpenCL engine provides an implementation of the Accelerator interface found in \texttt{ACC}, thus allowing access to a GPU or a CPU-based OpenCL platform, then widely used in the development of the CPI optimized application. 
\end{enumerate}

\begin{figure}[tb] 
\centering 
\includegraphics[width=0.7\linewidth]{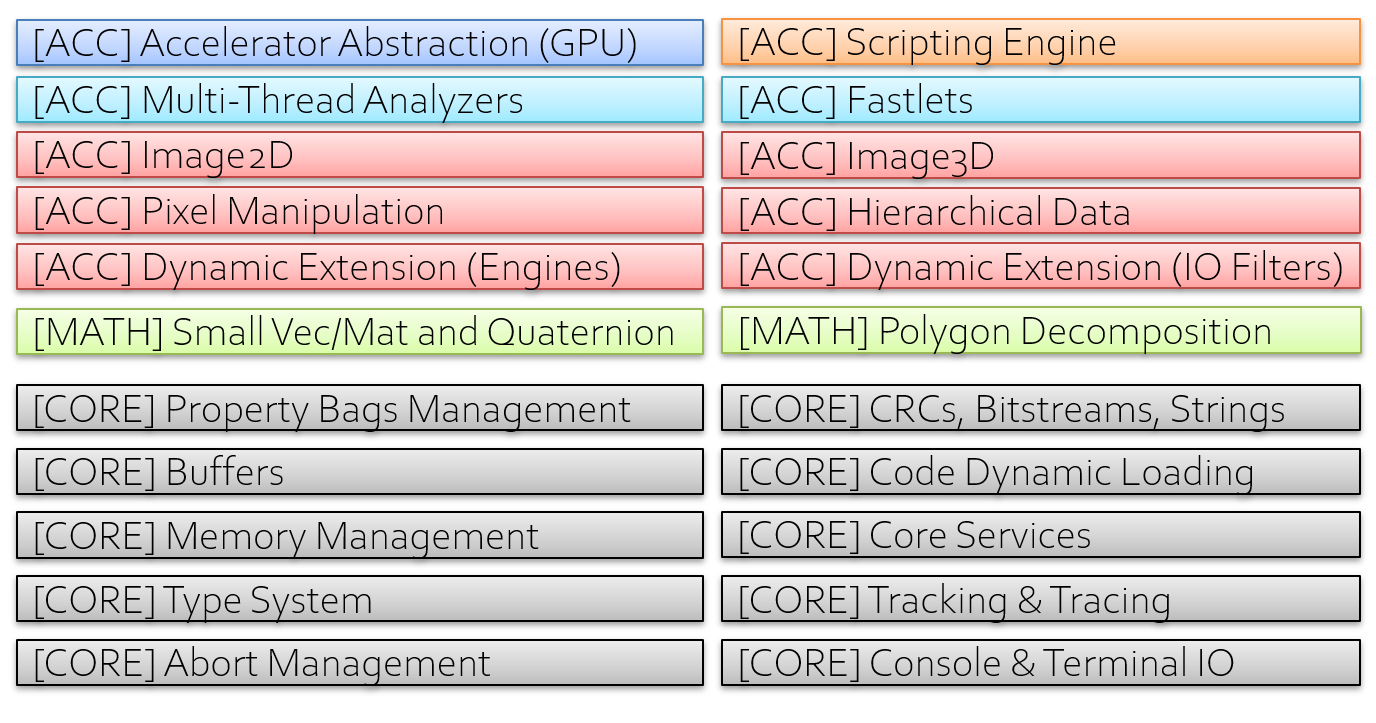}
\caption{Map of lower-level SDK functionalities.}
\label{fig:SDKoverview} 
\end{figure}

The diagram in Fig.~\ref{fig:SDKoverview} shows an overview of the basic libraries and their respective functionalities, detailed in Section \ref{subsubsec321}.

%%==================================%%
\subsubsection{SDK Core Functionalities}\label{subsubsec321}
In this section, we provide a high-level description of the core functionalities offered by the aforementioned SDK, implicitly exploited in designing and developing the optimized SW version of CPI application. 

\paragraph{Memory Management} The framework is designed by considering the typical memory management practice (including the type of allocations) developers face when writing C/C++ programs, i.e. static allocations in the data segment, allocation of dynamic memory (heap memory), and automatic allocation, implying the use of the stack.
To this end, the framework provides a set of abstractions of the allocation/de-allocation functionalities, which proved to be relevant in handling the large volumes of data required by the CPI approach.

\paragraph{Dynamic Loading of Code}
The use of Dynamic-Link Library (DLL) allows the implementation of extensible applications by using the concept of a plugin. There are specialized classes responsible for interfacing with the operating system for using the capability of opening DLLs, searching for specific symbols (mostly representing functions), and creating a sort of binding between the application and the code these functions contain. This allows the application to be extended with functionality on the fly, assuming that the overall architecture of the application is appropriately designed through the use of specific concept modelling.

\paragraph{Type System}
The SDK has its internal type-system to allow easy management of data and their conversion functions to one another and any suitable string representation. The type-system is used in any part of the framework, mostly for attributes management, hierarchical data handling and generic serialization/deserialization to/from buffers and strings.

\paragraph{Abort Management}
The SDK allows to manage eventual issues in software application behaviour coming from 
e.g. incomplete input validation code, incorrect memory management in terms of allocation and de-allocation, error in the implementation, and Hardware (HW) errors, that can cause
a degraded SW behaviour. To manage these issues, always present in the SW lifecycle as a result of the development, testing, and validation process, the framework introduces the concept of “degraded execution mode” and “predictive bug management” through the \textit{Abort Management} sub-system.

\begin{figure}[tb] 
\centering 
\includegraphics[width=0.7\linewidth]{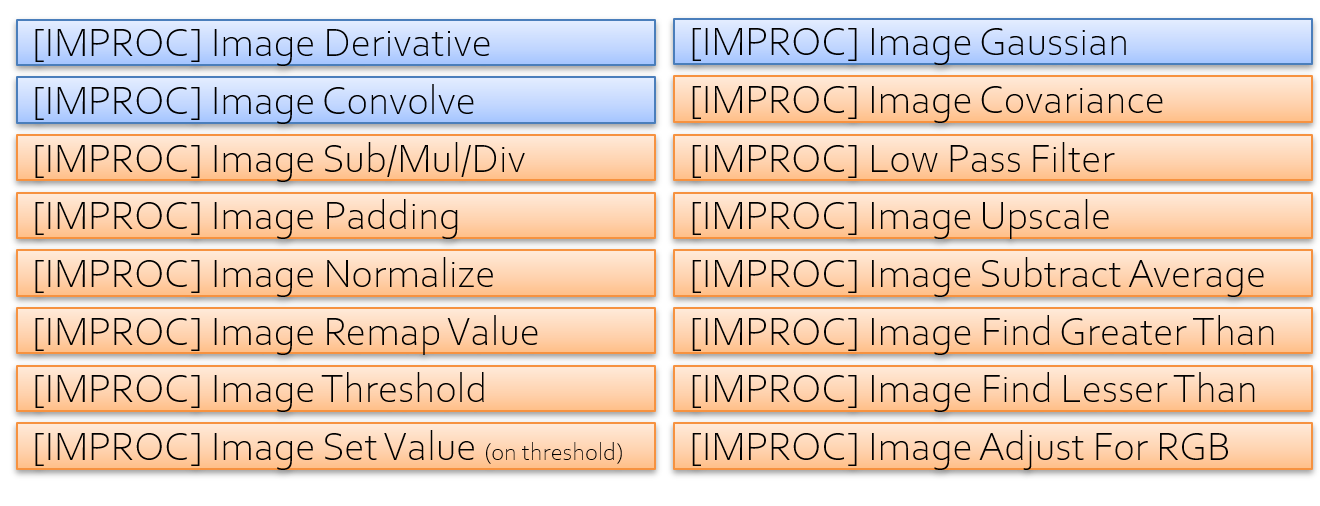}
\caption{SDK functionalities specifically devoted to Image Processing (\textit{improc}).}
\label{fig:SDK_improc} 
\end{figure} 

%%==================================%%
\subsubsection{SDK CPI-specific Functionalities}\label{subsubsec322}
The current section provides a high-level description of some relevant functionalities included in the SDK and strictly related to activities of SW development and optimization customized for the CPI application.
\paragraph{Image Processing}
The \textit{improc} library provides functionalities and classes related to multi-dimensional image processing, spanning from low-level pixel-based operations (e.g. thresholding, subtraction, multiplication etc) to more elaborated filtering (e.g. convolution, derivative, gaussian, etc), as resumed in Fig.~\ref{fig:SDK_improc}. The following two types of image data made available within the SDK were largely adopted during the CPI algorithm refactoring to represent bi-dimensional and 3-dimensional image data, respectively.

\texttt{Image 2D - }
The \texttt{CImage2D} class of \texttt{ACC} is an abstraction of an ordered sequence of pixel of the same type laid out in main memory or on file system through the memory-mapping API of the operating system. \texttt{ACC} provides access classes (readers and writers) for performing operations on a \texttt{CImage2D} instance.

\texttt{Image 3D - }
The \texttt{CImage3D} class of \texttt{ACC} is an abstraction of a vector of \texttt{CImage2D} of pixels of the same type laid out as determined by the \texttt{CImage2D} instances inside it. \texttt{ACC} provides access classes (readers and writers) for performing operations on a \texttt{CImage3D} instance. The pixel of a \texttt{CImage3D} is a vector of \texttt{D} components of the same type, each representing the pixel in the corresponding layer.

\begin{figure}[tb] 
\centering 
\includegraphics[width=0.8\linewidth]{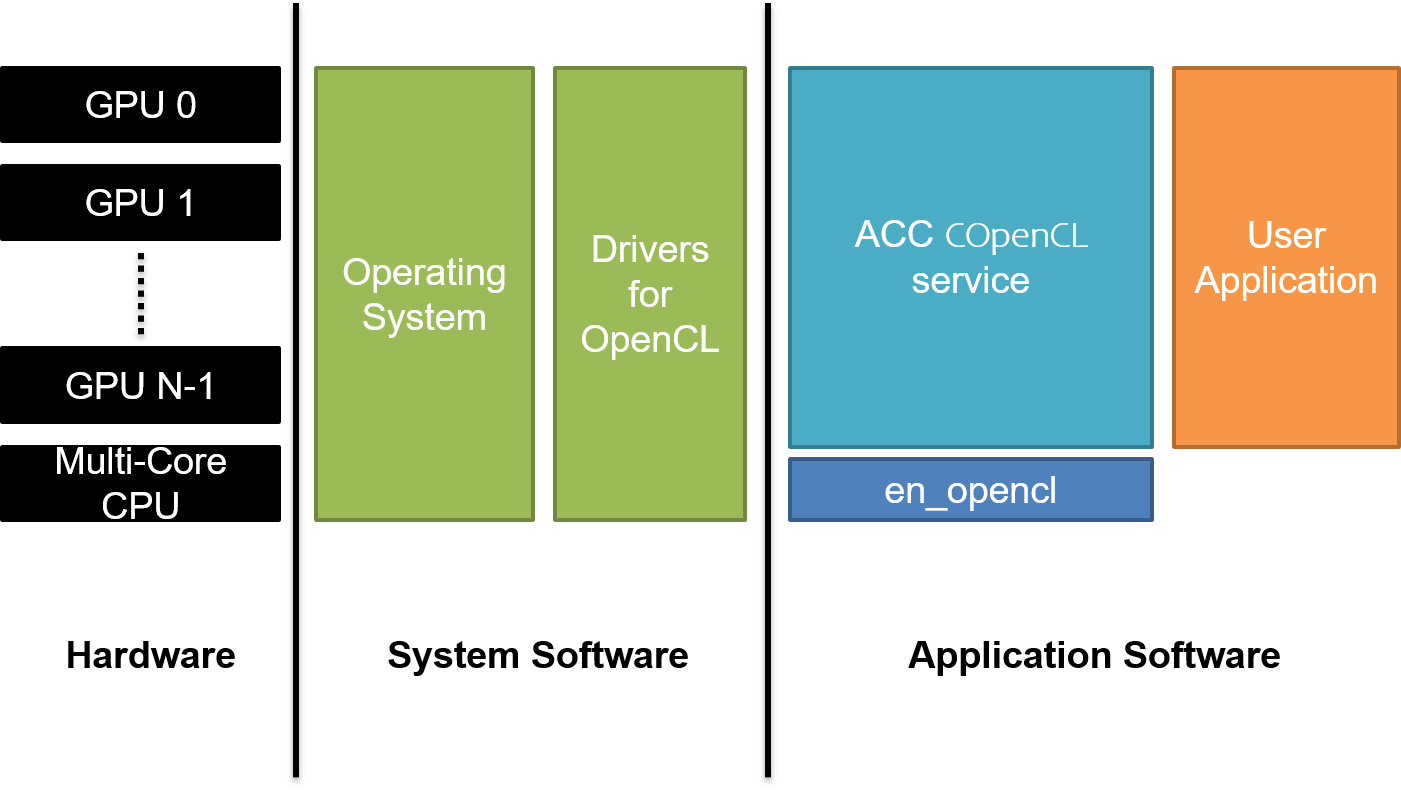}
\caption{Multi-core acceleration managed by the SDK.}
\label{fig:SDK_accel1} 
\end{figure}

\paragraph{Acceleration Abstraction: multi-thread analysers and GPU}
The SDK infrastructure related to Abstract Acceleration provides support for multi-CPU, multi-GPU and multi-core architectures, as sketched in Fig.~\ref{fig:SDK_accel1}. A central role in this context is covered by the SDK-related concept of \textit{Analyser}.

\begin{figure}[tb] 
\centering 
\includegraphics[width=0.8\linewidth]{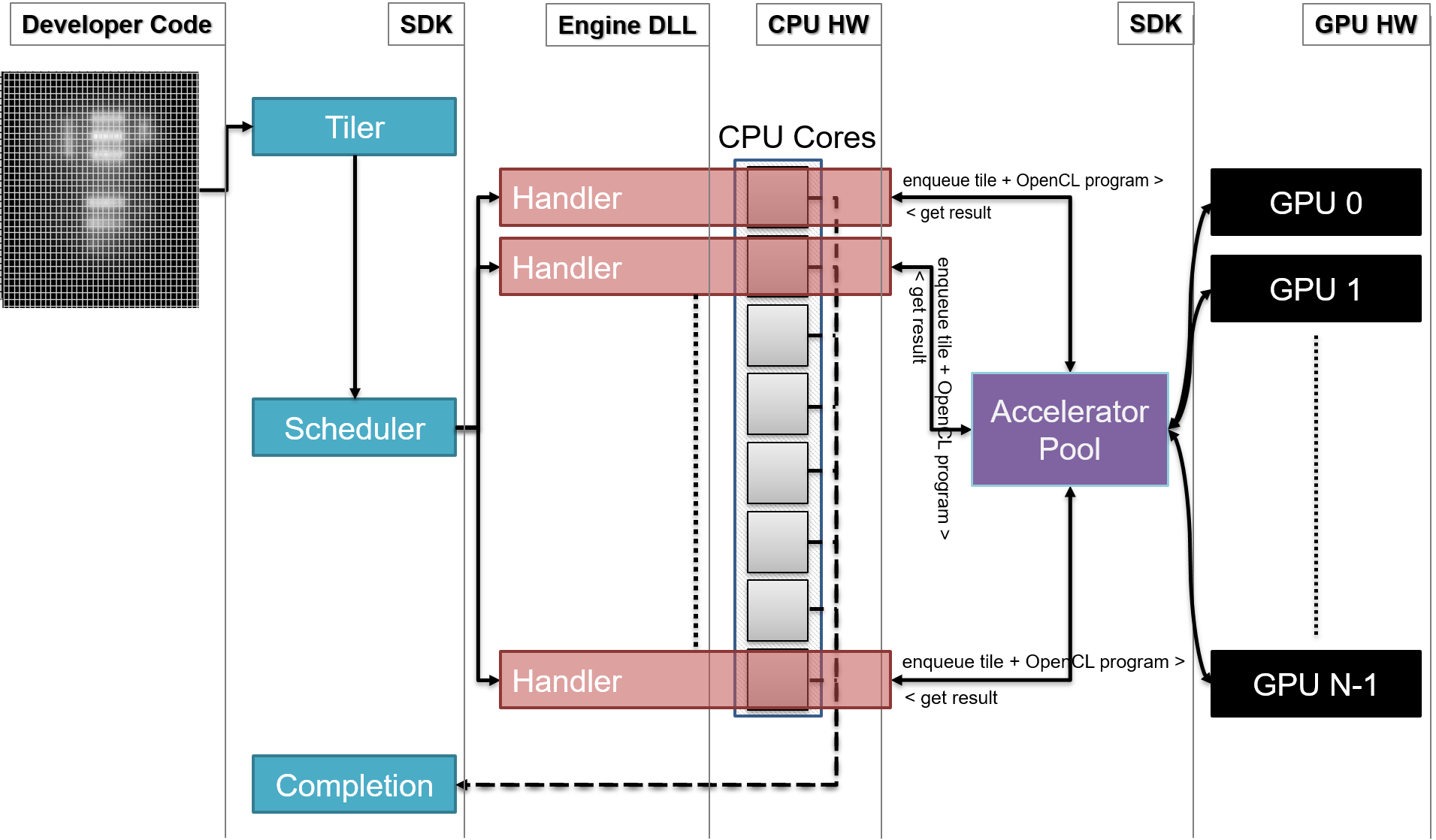}
\caption{The Analyzer mechanics (multi GPU).}
\label{fig:SDK_AnalyzerGPU} 
\end{figure} 

\begin{figure}[tb] 
\centering 
\includegraphics[width=0.8\linewidth]{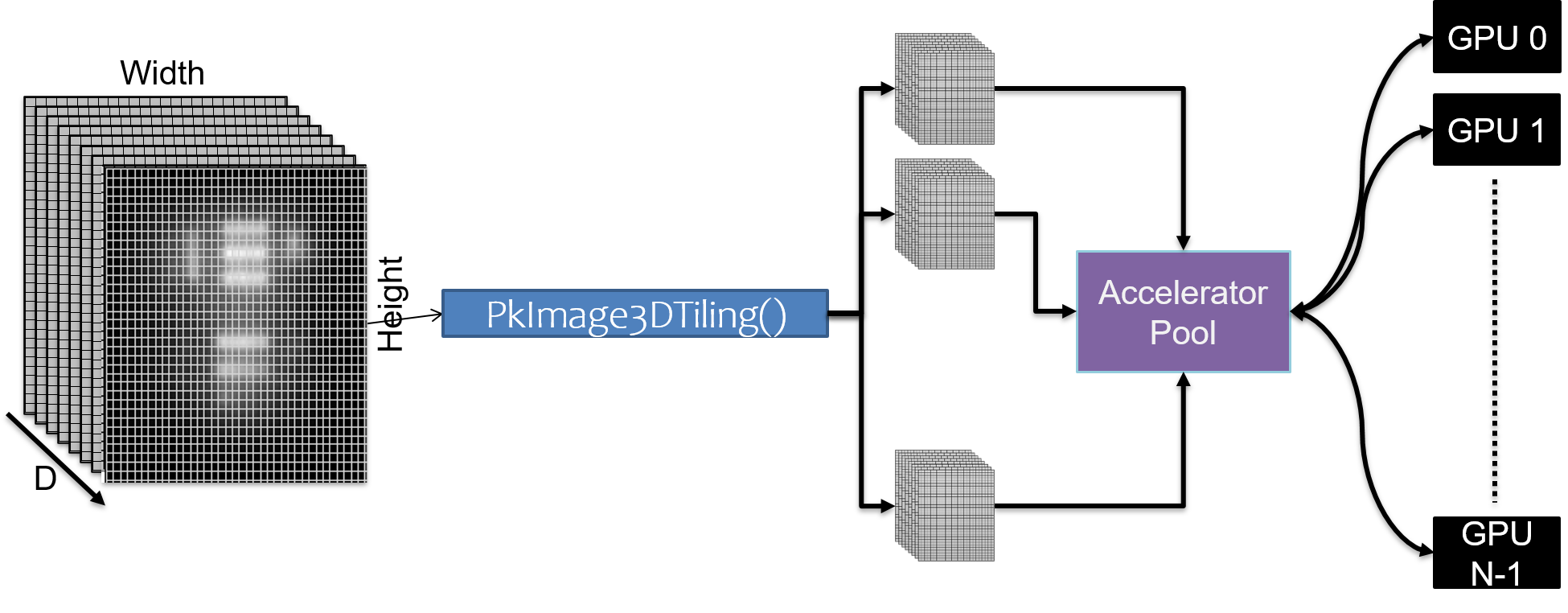}
\caption{The Analyzer mechanics applied to an instantiated \texttt{CImage3D}.}
\label{fig:SDK_AnalyzerCImage3D} 
\end{figure}

The \texttt{Analyser} is an entity of the framework that implements a \textit{map/reduce} paradigm for performing computations on images of any size, preferably large images, where the paradigm can offer a relevant improvement in terms of computational performances. More specifically, the framework entities related to \texttt{Analysers} will get the input image, divide it in tiles and then schedule a thread for each tile always limiting the number of concurrent threads to the hardware-concurrency of the host system.
The processing on the specified element can be activated on the whole image, or onto a set of sub-images of the main image. Consider Fig.~\ref{fig:SDK_AnalyzerGPU} for a schematic of the Analyzer mechanics augmented by the use of GPU, while Fig.~\ref{fig:SDK_AnalyzerCImage3D} shows the case of accelerator applied to a \texttt{CImage3D} object.

\begin{figure}[tb] 
\centering 
\includegraphics[width=0.5\linewidth]{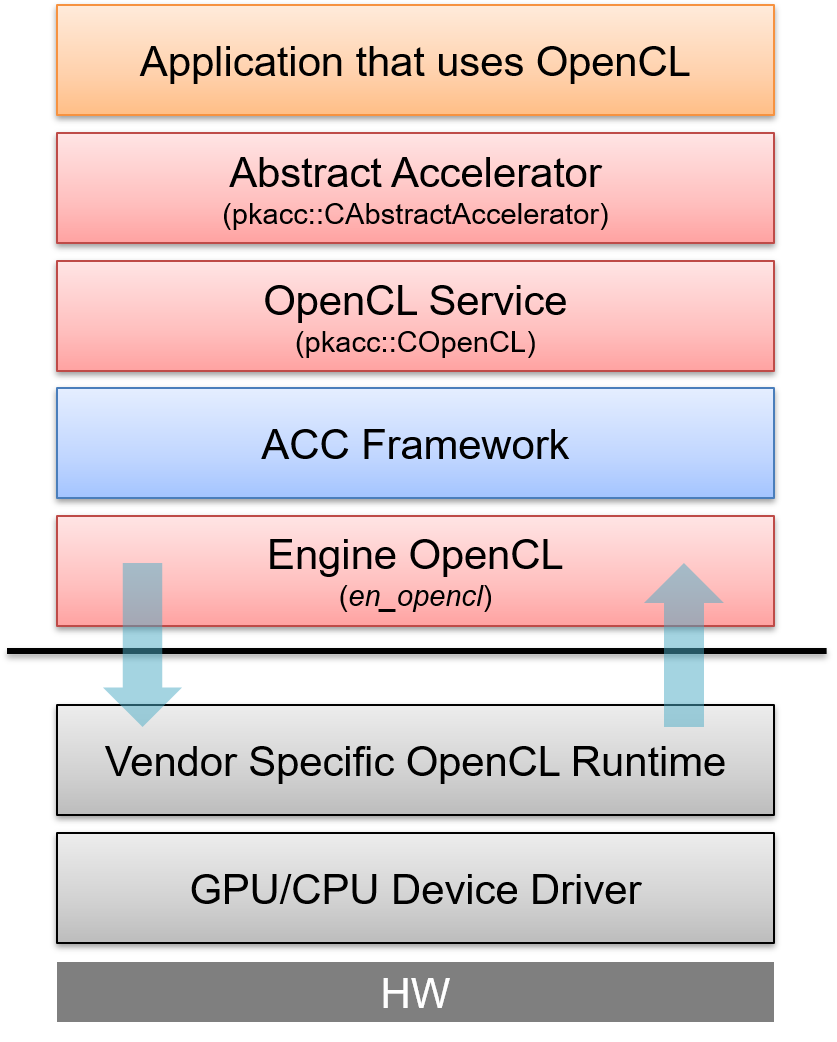}
\caption{OpenCL specialization of ACC Abstract Accelerator.}
\label{fig:SDK_OpenCL} 
\end{figure}

\texttt{OpenCL (Open Computing Language)} is an open standard for writing programs that execute across heterogeneous platforms consisting of central processing units (CPUs), graphics processing units (GPUs), and other processors or hardware accelerators from different vendors. In this context, particular attention has been provided in empowering the CPI-relate framework with OpenCL-based functionalities in order to ensure cross-vendor portability and cross-device compatibility.
To this end, the \texttt{ACC} library provides the capability, through the \texttt{pkacc::CAbstractAccelerator} related infrastructure, for building acceleration engines based on OpenCL.
As presented in Fig.~\ref{fig:SDK_OpenCL}, the \texttt{COpenCL} Service provides instances of \texttt{CAbstractAccelerator}, one for each detected OpenCL capable device found on the system and presented by the operating system and its device drivers.
The abstraction adapts entities of the SDK, such as  \texttt{Image2D},  \texttt{Image3D} and Buffer to what the OpenCL standard expects to exchange with the GPU or any supported accelerator. This is done concretely into the OpenCL engine named \texttt{en\_opencl}, which shall be built by linking the Vendor specific SDKs, and loaded by the runtime environment of the SDK framework.

%%==================================%%
%%    Results and Discussion        %%
%%==================================%%
\section{Results and Discussion}\label{sec4}
\subsection{CPI enhancement in GPU-based environment}\label{subsec41}
The results discussed here were aimed at optimising CPI computational performance by applying an algorithm refactoring focused on the implementation of customised OpenCL features for execution on parallel computing accelerators. 
A strategy based on a hybrid solution involving both CPU-based and GPU devices has been followed, implementing an OpenCL-based processing kernel. In detail, several features offered by our SDK have been exploited to develop programs that can be executed on heterogeneous platforms while being device-agnostic. In this context, the OpenCL standard ensures that the complexity of translating entities of the SDK (such as \texttt{CImage2D}, \texttt{CImage3D}, buffers, etc.) to what is expected to be exchanged with the accelerator device (e.g. a GPU) is left to the infrastructure Abstraction Layer, and this removes the need for the end user to explicitly manage the processing devices.

The considered algorithmic procedure consists of the following two main steps to be optimized: (i) calculation of the Cross-Correlation matrix between two Regions-of-Interest (RoI), and 
 (ii) the Refocusing of the image for a specific refocusing plane. 
For benchmarking purposes, the selected use case is considered to analyze and cross-correlate two images, each of 256x256 pixels.

\paragraph{Cross-correlation Optimization}
In the original version of the prototypal code, considered as a baseline for performance comparison, the files containing the frames/images of the input are read, bit-unpacking is performed, and the cross-correlation between the two ROIs is computed by matrix multiplication (Eq.~\eqref{eq:corrFunc}).
More specifically, the refactoring of this step relies on the fact that the input binary data are formatted as binary information (i.e. each pixel of the acquired images can assume only 0 or 1 value alternatively). In this scenario, a strategy based on bit-packing has been adopted, then implementing the cross-correlation multiplication as a bit-by-bit \texttt{AND} operation, thus exploiting capabilities of parallel processing (i.e. performing the same operation on multiple data points simultaneously).
In the optimized OpenCL solution, bit unpacking and cross-correlation calculation occur jointly in the same kernel by bitwise operations. Moreover, since the size of the cross-correlation matrix may not fit entirely into the memory of the specific device, the computation is divided into a dynamic number of blocks, depending on the memory available.

\paragraph{Refocusing Optimization}
Considering the refocusing step, the prototypal implementation include a 4D multilinear interpolation of the resulting cross-correlation matrix. The points of interpolation vary according to the selected refocusing plane and are usually computed images for multiple refocusing planes. Moreover, the prototypal implementation of the algorithm requires a relevant quantity of RAM ($\sim100$  GB) to run on 256x256 images, and is therefore unusable on more modest hardware.
In the optimized OpenCL solution, refocusing requires little more than the memory required to keep the same cross-correlation matrix and output files in RAM, as the interpolation coordinates are computed on-the-fly in the main refocusing kernel.
Again, for reasons similar to the cross-correlation calculation, the algorithm proceeds by blocks. The number of blocks is determined dynamically based on the memory available on the device used for acceleration, thus improving the achievable overall computing performances, based on the actual hardware available for processing.

%%==================================%%
\subsection{SW benchmarks}\label{subsec42}
In the current section, we present a comparison of the execution times of the algorithm in its original version, and in the optimized version (w\ OpenCL-based kernel). Since in both cases the algorithm consists of two independent parts (cross-correlation and refocusing), whose execution times scale based on different factors, the benchmarks have been reported separately for the two sections of the algorithm in both prototypal code and OpenCL. Note that because the data and the output data format is different in the two implementations, the times for writing the data to disk have been excluded from the comparison.

\begin{table}[h]
\caption{HW specifications of platform(s) used for benchmarks. \textit{Left:} Test Platform specs. \textit{Right:} Target Platform specs.  }\label{tabHWenv}%
\begin{tabular}{@{}lll@{}}
\toprule
& Test Platform & Target Platform \\
\midrule
OS     & Ubuntu 20.04 & Ubuntu 18.04  \\
CPU    & Intel(R) i7-9700k @3.60GHz & AMD Ryzen Threadripper PRO3955WX @3.90GHz\\
RAM    & 64 GB    & 256 GB \\
GPU    & NVIDIA RTX 2080Ti [11GB] & NVIDIA RTX4090
[24 GB] \\
\botrule
\end{tabular}
\end{table}

\begin{figure}[tb] 
\centering 
\includegraphics[width=0.8\linewidth]{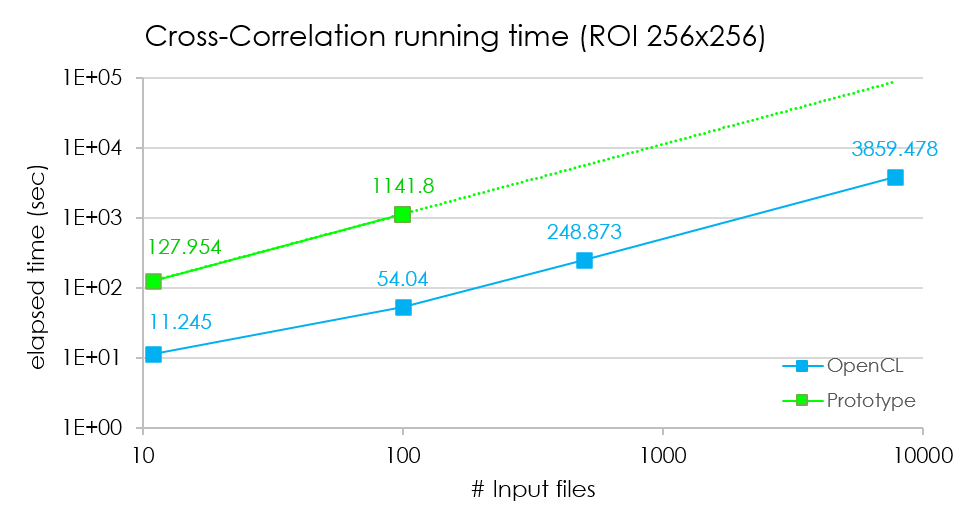}
\caption{CPI Cross Correlation: running time benchmark of prototypal implementation VS Open-CL optimization.}
\label{fig:benchCrossCorr} 
\end{figure}

\begin{figure}[tb] 
\centering 
\includegraphics[width=0.8\linewidth]{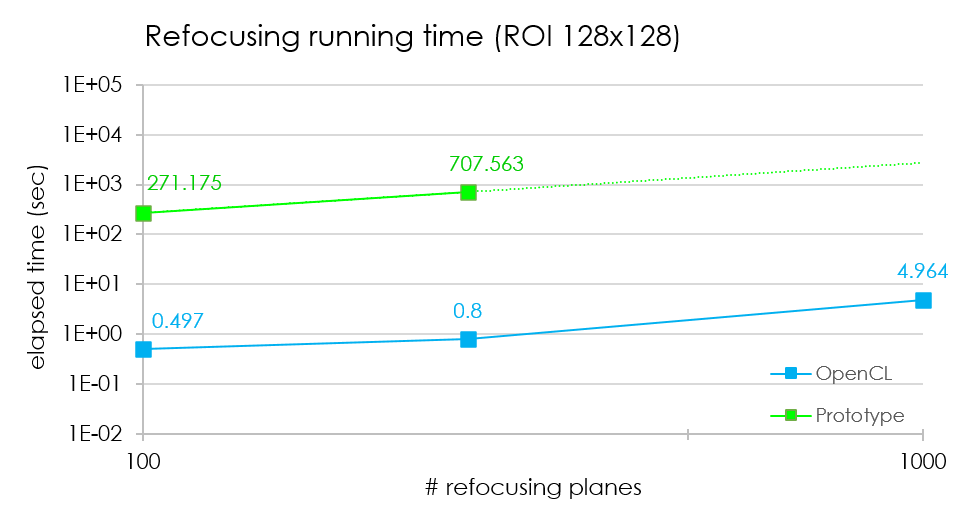}
\caption{CPI Refocusing: running time benchmark of Open-CL optimization (RoI: 128x128).}
\label{fig:benchRefoc128} 
\end{figure}

\begin{figure}[tb] 
\centering 
\includegraphics[width=0.8\linewidth]{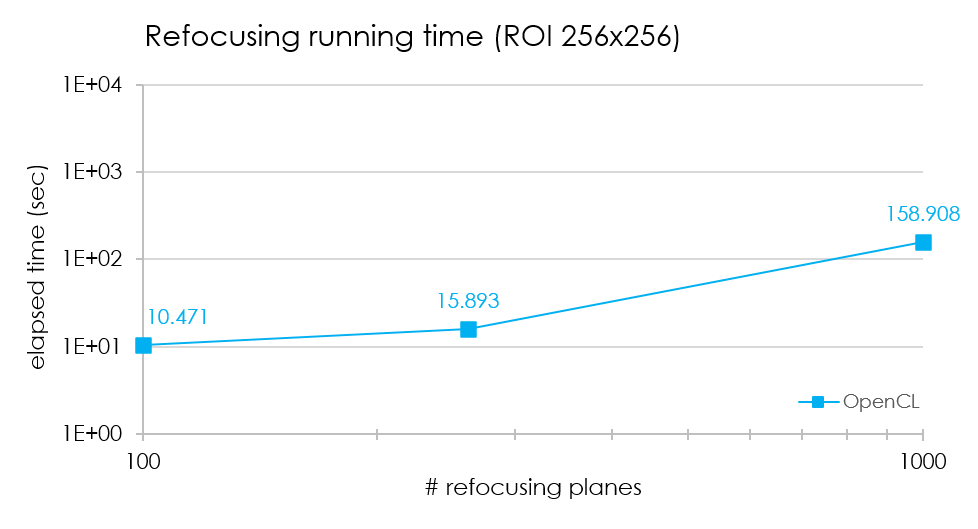}
\caption{CPI Refocusing: running time benchmark of Open-CL optimization (RoI: 256x256).}
\label{fig:benchRefoc256} 
\end{figure}

Relevant specifications of the HW test platform used for benchmarking are reported in Table \ref{tabHWenv}. In detail, results presented in Figs.~\ref{fig:benchCrossCorr}--\ref{fig:benchRefoc256} were obtained using the Test Platform (Table \ref{tabHWenv}). 

\paragraph{Cross-correlation}
Benchmarks for cross-correlation were performed on areas of interest of size 256x256. Since the execution time for cross-correlation scales according to the number of frames (and thus files) as input, benchmarks were performed using the following incremental number of files: 11, 100, 500, 7808 (size of the full dataset). Given the long run times, benchmarks have been omitted for prototypal runs for 256x256 images with 500 and 7808 files, but the extrapolation of the time elapsed in these cases was performed and shown in Fig.~\ref{fig:benchCrossCorr}. Results obtained in terms of average execution time are presented in graphical format in Fig.~\ref{fig:benchCrossCorr}.

\paragraph{Refocusing}
Benchmarks for refocusing were performed on both 128x128 and 256x256 images for OpenCL, but only 128x128 in the case of prototypal implementation, since the latter requires an amount of RAM memory not available on the workstation used. 
Since the execution time of the refocusing step is independent of the number of input files input, but depends only on the amount of output images to be produced (i.e., it varies according to the amount of refocusing planes desired), the benchmarks were performed with a fixed number of files (11 files) and with a variable number of refocusing planes: 100, 261, 1000. Results are presented for 128x128 and 256x256 RoI in Figs.~\ref{fig:benchRefoc128}-\ref{fig:benchRefoc256}, respectively.


\subsubsection{Performance VS GPU-memory}\label{subsubsec421}
As an additional experiment, we tested the OpenCL-based optimized version of the Cross-Correlation processing using different platforms mounting NVIDIA GPUs with increasing available memory. The activity described above is aimed at investigating the potential in achieving near-realtime computation when a processing device with a powerful GPU is available. In this case, the analysis is performed for cross-correlation only, being this step the most challenging in terms of computational times, and thus the potentially limiting factor for future perspective going towards the design of a near-real time CPI device.

\begin{figure}[tb] 
\centering 
\includegraphics[width=0.7\linewidth]{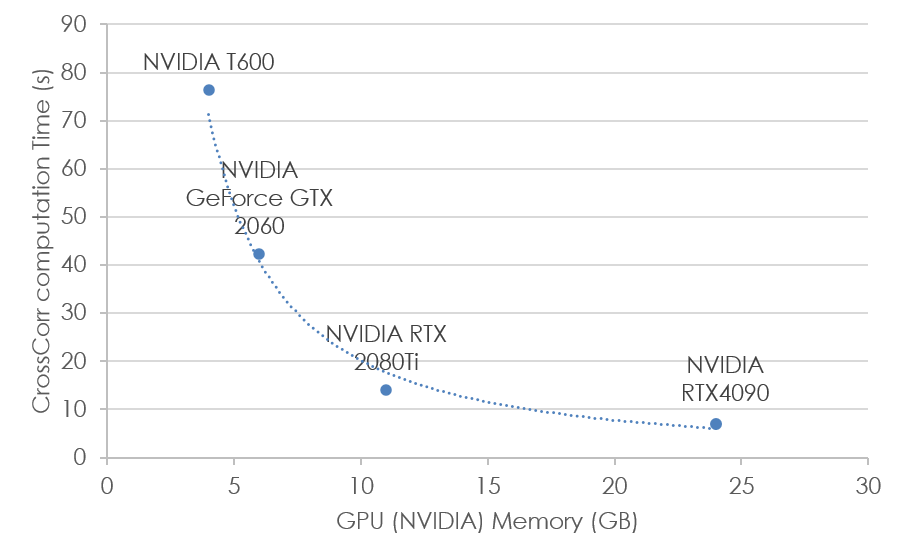}
\caption{Optimized Performance achieved in CrossCorr computation versus GPU device memory. Experiment performed using the following configuration of parameters: RoI 256x256, input data subset made of $\sim$ 10 thousand binary frames.}
\label{fig:benchVsGPUmemory} 
\end{figure}

Figure \ref{fig:benchVsGPUmemory} shows the total amount of time for Cross-Correlation processing (expressed in seconds) as a function of the nominal GPU memory made available by the processing device. Benchmarks reported here were performed using a significant subset of the input dataset (see Section \ref{sec2}), that can be considered representative of a typical use case for CPI application.

%%=======================%%
%%    Con lusions        %%
%%=======================%%
\section{Conclusion}\label{concl}

Achievements presented in previous sections demonstrate the actual ability to improve the CPI processing capabilities when moving from the baseline prototypal version (entry point of the activity) to the final GPU-optimized implementation. 
Benchmarks presented in Section 4 show that the optimized OpenCL-based implementation provides relevant improvements in the computational time elapsed for the target CPI algorithm, quantified as 10x to 20x for the Cross-correlation step, up to the impressive result of \textgreater 500x for the Refocusing part. In addition, it has been proved that the proposed implementation is able to scale performance according to the HW available, thus leaving open the possibility of including a better performing HW (i.e. GPU with larger memory) in order to achieve the desired computing performances.

\backmatter

\bmhead{Data availability statement}
All data are available upon request to the corresponding authors.

\bmhead{Acknowledgements}
Activity funded within the frame of Qu3D Project, supported by the Italian Istituto Nazionale di Fisica Nucleare, the Swiss National Science Foundation (grant 20QT21 187716 “Quantum 3D Imaging at high speed and high resolution”), the Greek General Secretariat for Research and Technology, the Czech Ministry of Education, Youth and Sports, under the QuantERA programme, which has received funding
from the European Union’s Horizon 2020 research and innovation programme.

\bibliography{bibliography.bib}


%\bibliography{bibliography.bib}% common bib file

\end{document}